# Quantum corrections to conductivity for semiconductors with various structures.

**S. A. Alavi, A. Tatar**
**Department of Physics, Sabzevar Tarbiat Moallem university,**
**P. O. Box 397, Sabzevar, Iran**
Email : alavi@sttu.ac.ir,

***Abstract*-We study the magnetic field dependences of the conductivity in heavily doped, strongly disordered 2D quantum well structures within wide conductivity and temperature ranges. We show that the exact analytical expression derived in our previous paper [1], is in better agreement than the existing equation i.e. Hikami(et.al.,) expression [2,3], with the experimental data even in low magnetic field for which the diffusion approximation is valid. On the other hand from theoretical point of view we observe that our equation is also rich because it establishes a strong relationship between quantum corrections to the conductivity and the quantum symmetry $Su_q(2)$. It is shown that the quantum corrections to the conductivity is the trace of Green function made by a generator of $Su_q(2)$ algebra. Using this fact we show that the quantum corrections to the conductivity can be expressed as a sum of an infinite number of Feynman diagrams.**

Keywords: magnetoresistance; Quantum corrections to conductivity; heterostructures; Feynman diagrams; Harper Hamiltonian; Quantum groups.
PACS: 71. 23. –k , 73. 43. Qt

## I. INTRODUCTION

Quantum transport phenomena and magnetotransport in a two dimensional electron gas (2DEG) have been subject of interest for scores of years. The quantum corrections to Drude conductivity in disordered metals and doped semiconductors has been most intensively studied for three decades. The quantum corrections to conductivity arises from interference of electron waves propagating in opposite directions along closed paths. The negative magnetoresistance induced by the suppression of the quantum interference correction by magnetic field is a manifestation of weak localization. It gives a simple analytical expression for quantum corrections to conductivity. An approximate expression for magnetic field dependence of magnetoresistance has been given by Hikami et.al., [2] and Wittman et.al.,[3]:

$$\Delta\sigma(B) = aG_0 H\left(\frac{\tau}{\tau_\phi}, \frac{B}{B_{tr}}\right),$$
$$H(x,y) = \Psi\left(\frac{1}{2}+\frac{x}{y}\right) - \Psi\left(\frac{1}{2}+\frac{1}{y}\right) - Ln(x), \quad (1)$$
$$B_{tr} = \hbar c/2el^2, \qquad G_0 = e^2/2\pi^2\hbar, \qquad \gamma = \tau/\tau_\phi .$$

where $\Psi(x), l$ , $\tau$ , and $\tau_\phi$ are digamma function, mean free path, mean free time and the phase breaking time respectively. The value of prefactor $a$ is equal to unity in non-interacting case. "(1)" has been used by Physicists to analyze experimental data and determine the temperature dependence of phase breaking time or length through the fitting of experimental curves [3,4,5]. A more exact expression has been derived by Alavi and Rouhani using quantum groups [1]:

$$\Delta\sigma(B) = -\sqrt{\frac{2}{\pi}}\alpha G_0 \sum_{N=3}^{\infty} e^{-\frac{N}{2\beta}} \{\cos^N\left(\frac{\left(\frac{B}{4\alpha B_{tr}}\right)}{\sinh\left(\frac{BN}{4\alpha B_{tr}}\right)}\right) \times \frac{\left(\frac{B}{4\alpha B_{tr}}\right)}{\sinh\left(\frac{BN}{4\alpha B_{tr}}\right)} - \frac{1}{N}\cos^N\left(\frac{1}{N}\right)\}, \qquad (2)$$

Here $\alpha = l^2$ , $\beta = ll_\phi$ and N is the number of collisions.

In this paper we study the magnetic field dependence of the conductivity in heavily doped**,** strongly disordered 2D quantum well structures within wide conductivity and temperature ranges**.** We show that expression (2) is more exact than expression (1), even in low magnetic field for which the diffusion approximation is valid.

## II. DATA ANALYSIS.

In this section we study the magnetic field dependence of $\Delta\sigma / G_0$ using "(1)" and "(2)" through the fitting of experimental data of heterostructure GaAs\InGaAs\GaAs in different conditions. In " (2)", $\alpha$ and $\beta$ and in "(1)" $a$ and $\gamma$ have been used as fitting parameters. In " (2)", N ranges from 3 to 900. In figure. (1), $\Delta\sigma / G_0$ is plotted against magnetic field for above mentioned heterostructure with $80\ \overset{0}{A}\ In_{0.2}Ga_{0.8}As$ as quantum well, $\delta - doped$ by Sn, for three temperatures. The gate voltage is $V_g = -1.8V$ and $k_f l$ = 17.9. As it is clear from figure (1), "(2)" is in better agreement than "(1)" with experimental data, specially in low temperatures where the quantum correction manifest itself more clear.

In figures (2) and (3), $\Delta\sigma/G_0$ is plotted against magnetic field B for above mentioned heterostructure with $80\ \overset{0}{A}\ In_{0.2}Ga_{0.8}As$ as quantum well and *heavily* $\delta - doped$ by Sn in the center. figure. (2) shows the result for the case of strongly disordered media with $V_g = -3.4V$ and $n = 6.5 \times 10^{11}\ cm^{-2}$ (n is the density of electron ), and figure (3) shows the result for the case of strongly disordered media with $V_g = -3.7V$ and $n = 5.2 \times 10^{11} cm^{-2}$ in two different temperatures.

As is observed from figs. (1)- (3), "(2)" is in better agreement than "(1)" with experimental data and "(1)" shows much more deviation in low temperatures, even in low magnetic field, see fig. (1).

## III. MAGNETORESISTANCE AND QUANTUM GROUPS

Let us consider a spinless electron on a two-dimensional lattice and submitted to a uniform magnetic field along the z-direction and perpendicular to the plane of motion. The system is not invariant under translations but there is an invariance under the so-called magnetic translation operators $w(\vec{a})$. The Harper model Hamiltonian in the absence of impurity is [6].

$$H_0 = \sum_{|a|=1} w(\vec{a}) = w(\vec{a}) + w(\vec{b}) + w(-\vec{a}) + w(-\vec{b}), \qquad (3)$$

where

$$w(\vec{a}) = \exp\left(\frac{i}{\hbar}\vec{a}\cdot\vec{k}\right), \qquad (4)$$

$$\vec{k} = \vec{p} + e\vec{A} + e\vec{r}\times\vec{B}. \qquad (5)$$

Here $\vec{a}$ and $\vec{b}$ are two perpendicular unit vectors which build the lattice unit cell, $|\vec{a}| = |\vec{b}| = 1$.

We can treat the impurity in the Harper model as a perturbation[7], in the sense that the Hamiltonian can be subdivided as

$$H = H_0 + V \qquad (6)$$

where $V$ is the potential created by impurities.

The probability distribution of closed random paths is given by [8]:

$$P(A,N) \equiv P_N(A) = \frac{2\pi N}{4^{N+1}}\frac{1}{\sqrt{2\pi}}\int_{-\infty}^{+\infty} e^{-iax}Tr\left(H_0^N(x)\right)dx, \qquad (7)$$

where $a = \frac{A}{N}$ is the renormalized area, A is the algebraic area and N is the number of collisions. Using the eigenvalues of the operator $j_y$ of the $Su_q(2)$ algebra derived in [9], It is shown that[9]:

$$Tr\left(H_0^N\right) = \frac{4^{N+1}}{2\pi N}\frac{\frac{x}{4}}{\sinh\left(\frac{x}{4}\right)}\cos^N\left(\frac{\frac{x}{4}}{N\sinh\left(\frac{x}{4}\right)}\right), \qquad (8)$$

where $x = \gamma N$, $\gamma = 2\pi\frac{\Phi}{\Phi_0}$ . $\Phi$ and $\Phi_0$ are the magnetic flux through the unit cell and the quantum of flux, respectively. In order to give an expression for conductivity correction in a magnetic field, we introduce the distribution of closed random paths of area $w_N(A)$ [10] such that $w_N(A)dA$ gives the probability density of return after N collisions following a trajectory which enclosed the area in the range (A,A + dA) , this gives [1]:

$$\delta\sigma(b) = -2\pi G_0\{\alpha\sum_{N=3}^{\infty}\int_{-\infty}^{+\infty} dA e^{-\frac{N}{2\beta}} w_N(A)\cos\left(\frac{(1+\gamma)^2 bA}{\alpha}\right)\}, \qquad (9)$$

where

$$w_N(A) = \frac{1}{\pi N}P_N(A), \qquad (10)$$

$$b = \frac{B}{(1+\gamma)^2 B_{tr}} . \qquad (11)$$

It is worth mentioning that "(2)" can be derived using "(7)", "(8)", "(9)" and "(10)".

On the other hand, Quantum algebras or q-algebras are the q-deformation of the ordinary Lie algebras [11]. Our argument is based on the quantum algebra $Su_q(2)$. The generators of the $Su_q(2)$ algebra satisfy the commutation relations:

$$[j_3, j_\pm] = \pm j_\pm , \tag{12}$$

$$[j_+, j_-] = \frac{q^{2j_3} - q^{-2j_3}}{q - q^{-1}}. \tag{13}$$

One can show that the following combinations of the magnetic translations:

$$j_+ = \frac{w(\vec{a}) + w(\vec{b})}{q - q^{-1}}, \tag{14}$$

$$j_- = -\frac{w(-\vec{a}) + w(-\vec{b})}{q - q^{-1}}, \tag{15}$$

satisfy the $Su_q(2)$ commutation relations "(12,13)". q is the parameter of deformation and is related to the magnetic field through the following relation :

$$q = \exp(i\frac{e}{\hbar}\vec{B}\cdot(\vec{a}\times\vec{b})) = e^{2\pi i\frac{\Phi}{\Phi_0}} . \tag{16}$$

From "(3)", "(14)" and "(15)" we obtain:

$$H_0 = (q - q^{-1})(j_+ - j_-) = 2i(q - q^{-1})j_y . \tag{17}$$

From "(7)","(9)" and" (10)" we have:

$$\delta\sigma(B) = -\sqrt{\frac{2}{\pi}}G_0\alpha\sum_{N=1}^{\infty} e^{-N/N_0} Tr(H_0^N(x)). \tag{18}$$

Untill now in this section we have briefly reviewed some of the previous results presented in [1]. Now we can write "(18)" as follows:

$$\delta\sigma(B) = -\sqrt{\frac{2}{\pi}}G_0 e^{\frac{1}{N_0}}\alpha\, Tr\left(\frac{1}{E - H_0}\right). \tag{19}$$

Here $H_0$ is the Harper Hamiltonian and is related to the generators of $Su_q(2)$ algebra through "(17)". By adding a small imaginary value $\pm i\varepsilon$ to the denominator in the expression (19) we see that the expression in the bracket is nothing but the advanced and retarded Green functions. Now, we introduce the operator T as follows:

$$T = V + V\frac{1}{E - H_0 \pm i\varepsilon}T . \tag{20}$$

where V is the interaction potential. By iteration we have:

$$T = V + V\frac{1}{E - H_0 \pm i\varepsilon}V + V\frac{1}{E - H_0 \pm i\varepsilon}V\frac{1}{E - H_0 \pm i\varepsilon}V + \ldots . \tag{21}$$

This leads to :

$$G = G_{\pm}^{(0)} + G_{\pm}^{(0)}TG_{\pm}^{(0)} + G_{\pm}^{(0)}TG_{\pm}^{(0)}TG_{\pm}^{(0)} + \cdots \tag{22}$$

where $G_{\pm}^{(0)} = \frac{1}{E - H_0 \pm i\varepsilon}$ are the unperturbed advanced and retarded Green functions.

Using "(20)", "(22)" can be rewritten as follows:

$$G = G_{\pm}^{(0)} + G_{\pm}^{(0)}VG_{\pm}^{(0)} + G_{\pm}^{(0)}VG_{\pm}^{(0)}VG_{\pm}^{(0)} + \cdots \tag{23}$$

So "(19)" for the quantum corrections to the conductivity can be written as:

$$\delta\sigma(B) = -\sqrt{\frac{2}{\pi}}G_0 e^{\frac{1}{N_0}}\alpha\, Tr\left(G_{\pm}^{(0)} + G_{\pm}^{(0)}TG_{\pm}^{(0)} + \cdots\right),$$

Or

$$\delta\sigma(B) = -\sqrt{\frac{2}{\pi}}G_0 e^{\frac{1}{N_0}}\alpha\, Tr\left(G_{\pm}^{(0)} + G_{\pm}^{(0)}VG_{\pm}^{(0)} + \cdots\right), \tag{24}$$

We can rewrite " (24)" as follows:

$$\delta\sigma(B) = \left(\sum_{n=1}^{\infty}\delta\sigma^{(n)}(B)\right), \tag{25}$$

in which for instance the $\delta\sigma^{(n)}(B),\ n = 0,\ldots,4$ are as follows :

$$\begin{aligned}
\delta\sigma^{(0)}(B) &= -\sqrt{\frac{2}{\pi}}G_0 e^{\frac{1}{N_0}}\alpha\, Tr\left(G_{\pm}^{(0)}\right),\\
\delta\sigma^{(1)}(B) &= -\sqrt{\frac{2}{\pi}}G_0 e^{\frac{1}{N_0}}\alpha\, Tr\left(G_{\pm}^{(0)}VG_{\pm}^{(0)}\right),\\
\delta\sigma^{(2)}(B) &= -\sqrt{\frac{2}{\pi}}G_0 e^{\frac{1}{N_0}}\alpha\, Tr\left(G_{\pm}^{(0)}VG_{\pm}^{(0)}VG_{\pm}^{(0)}\right),\\
\delta\sigma^{(3)}(B) &= -\sqrt{\frac{2}{\pi}}G_0 e^{\frac{1}{N_0}}\alpha\, Tr\left(G_{\pm}^{(0)}VG_{\pm}^{(0)}VG_{\pm}^{(0)}VG_{\pm}^{(0)}\right)\\
\delta\sigma^{(4)}(B) &= -\sqrt{\frac{2}{\pi}}G_0 e^{\frac{1}{N_0}}\alpha\, Tr\left(G_{\pm}^{(0)}VG_{\pm}^{(0)}VG_{\pm}^{(0)}VG_{\pm}^{(0)}VG_{\pm}^{(0)}\right)
\end{aligned} \tag{26}$$

## IV. THE FEYNMAN DIAGRAMS

Equation (24) involves both single-particle Green's function i.e. $G_+^{(0)}VG_+^{(0)}$, $G_-^{(0)}VG_-^{(0)}$, … and two-particles correlation functions i.e. $G_+^{(0)}VG_-^{(0)}$, $G_-^{(0)}VG_+^{(0)}$ …, which we study now. The propagator (displayed in the momentum representation) depends parametrically on the impurity positions.

$$G(p,t;p',t') \equiv G(p,t;p',t';r_1,r_2,..r_N) \tag{27}$$

The impurity-averaged propagator can be obtained by averaging each term in the perturbative expansion of the propagator in terms of the impurity potential which has been expressed in terms of the individual impurity potentials

$$V(p-p') \equiv V(p,p',r_1,r_2,..r_N) = V_{imp}(p-p')\sum_{i=1}^{N} e^{-\frac{i}{\hbar}(p-p')\cdot r_i} \tag{28}$$

we have associated with each potential term the impurity phase factor [12].

$$\rho_{imp}(p-p';r_1,r_2,..r_N) = \sum_{i=1}^{N} \exp(-\frac{i}{\hbar}(p-p')\cdot r_i). \tag{29}$$

***a). Feynman diagrams for single-particle Green's function.***

Here we derive the Feynman diagrams for the retarded Green function $G_+^{(0)}$, the same method can be used to derive the Feynman diagrams for the advanced Green function $G_-^{(0)}$.

By averaging for the first-order advanced green function we have:

$$G_+^{(1)}(p,p',E) = G_+^{(1)}(p,E)\langle p|\hat{V}|p'\rangle G_+^{(1)}(p',E) \tag{30}$$

We get in the momentum representation

$$\left\langle G_+^{(0)}(p,p',E)\right\rangle = G_+^{(0)}(p,E)V_{imp}(p-p')G_+^{(0)}(p',E)\frac{1}{V}\left\langle \sum_{i=1}^{N} e^{-\frac{i}{\hbar}(p-p')\cdot r_i}\right\rangle \tag{31}$$

where we have used $V_{imp}$ instead of $V$. When $p \neq p'$, the oscillating exponents average to zero, and for $p = p'$, we get the number of impurity; i.e.,

$$\left\langle \sum_{i=1}^{N} e^{-\frac{i}{\hbar}(p-p')\cdot r_i}\right\rangle = N\delta_{p,p'} \tag{32}$$

Because there are N terms in the sum giving identical contributions. So for the first-order impurity-averaged propagator

$$\left\langle G_+^{(1)}(p,p',E)\right\rangle = n_i V_{imp}(p=0)\left[G_+^{(0)}(p,E)\right]^2 \delta_{p,p'}. \tag{33}$$

where $n_i$ is the impurity density. We can depict the factor in "33" :

$$G_+^{(1)}(p,E) = n_i V_{imp}(p=0)\left[G_+^{(0)}(p,E)\right]^2 \tag{34}$$

diagrammatically as :

$$G_+^{(1)}(p,E) = $$ 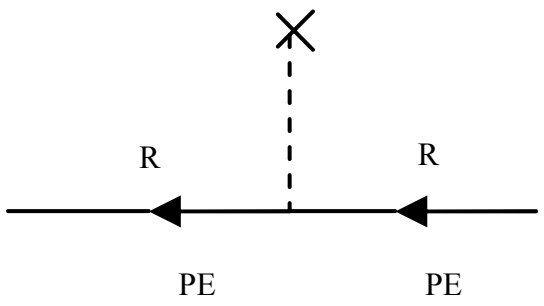

(35)

In which we have introduced :

$p$ 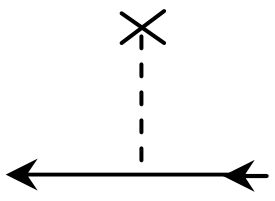 $p'$ $= n_i V_{imp}(p-p')$ (36)

Where the cross designates the impurity concentration, $n_i$, and the dashed line the Fourier transform of the impurity potential, $V_{imp}(p-p')$.

So the Feynman diagram (and rule) corresponding to the first order of quantum corrections to conductivity (QCC) is as follows:

$\delta\sigma^{(1)}(B) \leftrightarrow$ 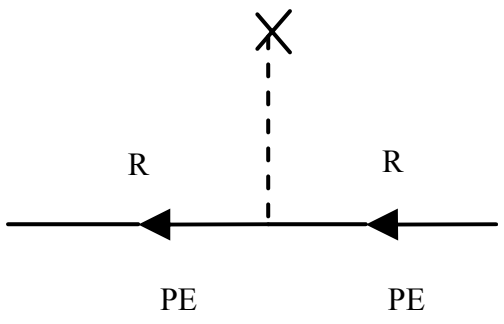

$$= -\sqrt{\frac{2}{\pi}} G_0 e^{\frac{1}{N_0}} \alpha \, Tr\left[n_i V_{imp}(p=0)\left[G_+^{(0)}(p,E)\right]^2\right] \tag{37}$$

For the second-order Green function

$G_+^{(2)}(p,p',E) = $ 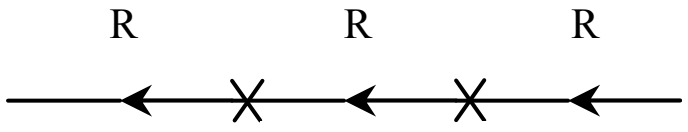

(38)

$pE \quad p''E \quad p'E$

We get, upon impurity averaging,

$$\left\langle G_+^{(2)}(p,p',E)\right\rangle = G_+^{(0)}(p,E)\sum_{p''} V_{imp}(p-p'')G_+^{(0)}(p'',E)V_{imp}(p''-p')$$
$$G_+^{(0)}(p',E)\frac{1}{V^2}\left\langle \sum_{i,j=1}^{N} e^{-\frac{i}{\hbar}(p-p'')\cdot r_i - \frac{i}{\hbar}(p''-p')\cdot r_j} \right\rangle \tag{39}$$

After averaging the sum over the impurity positions, for the contribution of scattering off different impurities we have :

$$\left\langle G_+^{(2)}(p,p',E)\right\rangle^{i\neq j} = \delta_{p,p'} n_i^2 \left[V_{imp}(p=0)\right]^2 \left[G_+^{(0)}(p,E)\right]^3$$
$$\equiv G_+^{(2)}(p,E)^{i\neq j}\delta_{p,p'} . \tag{40}$$

The diagrammatic representation of the prefactor is :

$G_+^{(2)}(p,E)^{i\neq j} =$

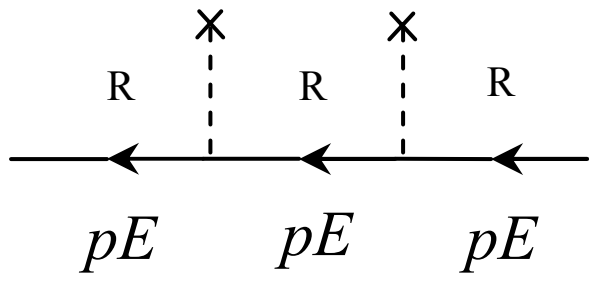

$$= \left[G_+^{(2)}(p,E)\right]^3 n_i^2 \left[V_{imp}(p=0)\right]^2 \tag{41}$$

Which can be kept or renormalized to zero.

The term corresponding to scattering off the same impurity $i = j$ gives the factor

$$\left\langle G_+^{(2)}(p,p',E)\right\rangle^{i=j} = n_i \delta_{p,p'} \left[G_+^{(0)}(p,E)\right]^2 \frac{1}{V}\sum_{p''} \left|V_{imp}(p-p'')\right|^2 G_+^{(0)}(p'',E)$$

$$\equiv G_+^{(2)}(p,E)^{i=j}\, \delta_{p,p'} \,. \tag{42}$$

The prefactor in this term has the diagrammatic representation

$G_+^{(2)}(p,E)^{i=j} =$ (43)

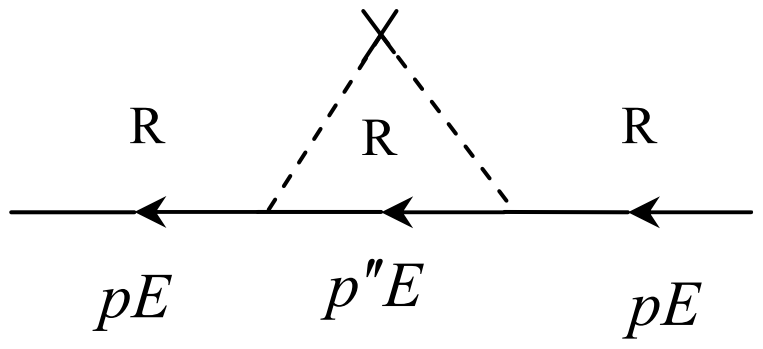

So the Feynman diagram (rule) corresponding to the second order QCC are :

$\delta\sigma^{(2)}(B) \leftrightarrow$

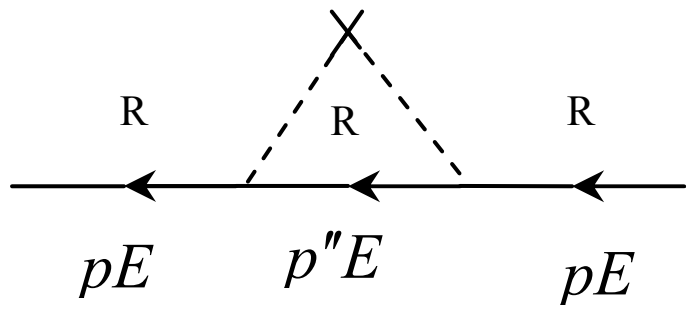

$$= -\sqrt{\frac{2}{\pi}}\, G_0 e^{\frac{1}{N_0}} \alpha Tr\left[ n_i \left[G_+^{(0)}(p,E)\right]^2 \int \frac{dp''}{(2\pi\hbar)^3} \left|V_{imp}(p-p'')\right| G_+^{(0)}(p'',E)\right]. \tag{44}$$

We note that in the continuum limit

$$\frac{1}{V}\sum_p \quad \rightarrow \int \frac{dp}{(2\pi\hbar)^3} \tag{45}$$

For the third-order diagram

$G_+^{(3)}(p,p',E) =$ (46)

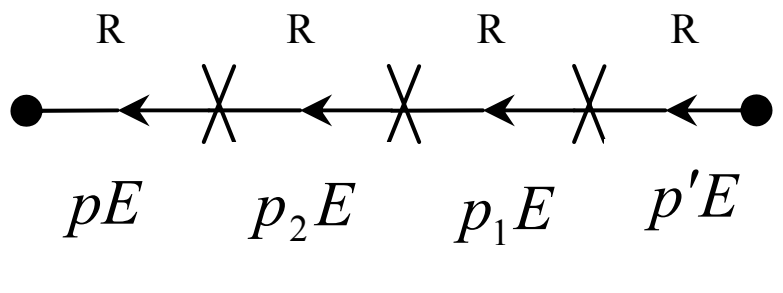

Which for triple scattering off the same impurity, $i = j = k$, gives the prefactor $N\delta_{p,p'}$ and thereby the following diagram:

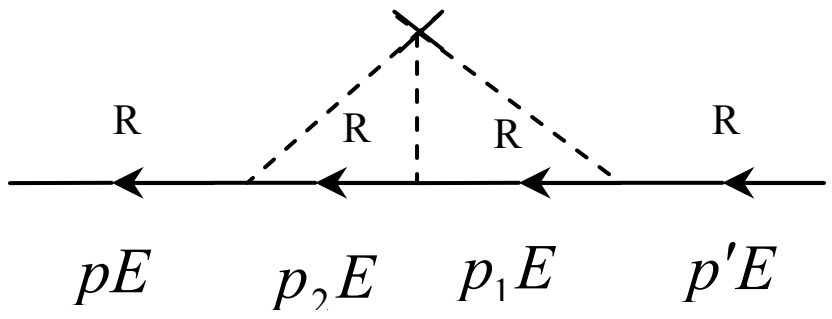

$$= n_i \left[G_+^{(0)}(p,E)\right]^2 \frac{1}{V^2} \sum_{p_1, p_2} V_{imp}(p - p_2) G_+^{(0)}(p_2, E) V_{imp}(p_2 - p_1) G_+^{(0)}(p_1, E) V_{imp}(p_1 - p) \tag{47}$$

where the three-leg represents the three impurity potential factors, and the cross, as before, The impurity concentration. After some calculation for the terms with double scattering off the same impurity i.e., the $i \neq j = k$, $i = j \neq k$ terms, we have:

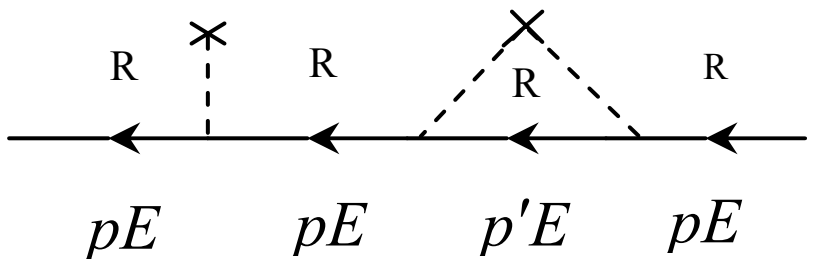

$$= n_i V_{imp}(p = 0) \left[G_+^{(0)}(p,E)\right]^3 n_i \frac{1}{V} \sum_{p'} \left|V_{imp}(p - p')\right|^2 G_+^{(0)}(p', E) \tag{48}$$

Which can be kept or renormalized to zero.

So the Feynman diagram corresponding to the third order QCC is as follows :

$\delta\sigma^{(3)}(B) \leftrightarrow$

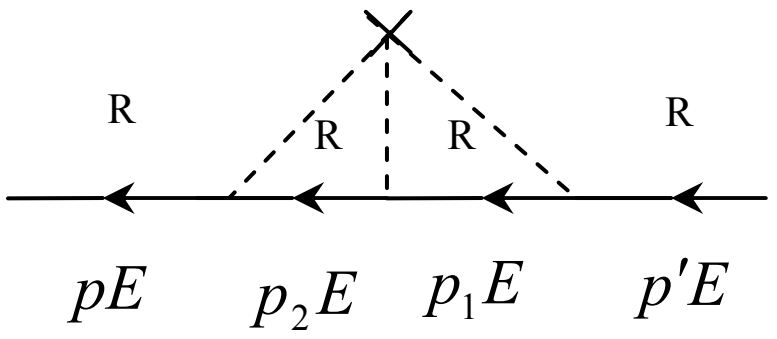

$$= -\sqrt{\frac{2}{\pi}} G_0 e^{\frac{1}{N_0}} \alpha\, Tr \left[ n_i \left[G_+^{(0)}(p,E)\right]^2 \int \frac{dp_1}{(2\pi\hbar)^3} \int \frac{dp_2}{(2\pi\hbar)^3} V_{imp}(p - p_2) G_+^{(0)}(p_2, E) V_{imp}(p_2 - p_1) G_+^{(0)}(p_1, E) V_{imp}(p_1 - p) \right] \tag{49}$$

The fourth-order diagram is

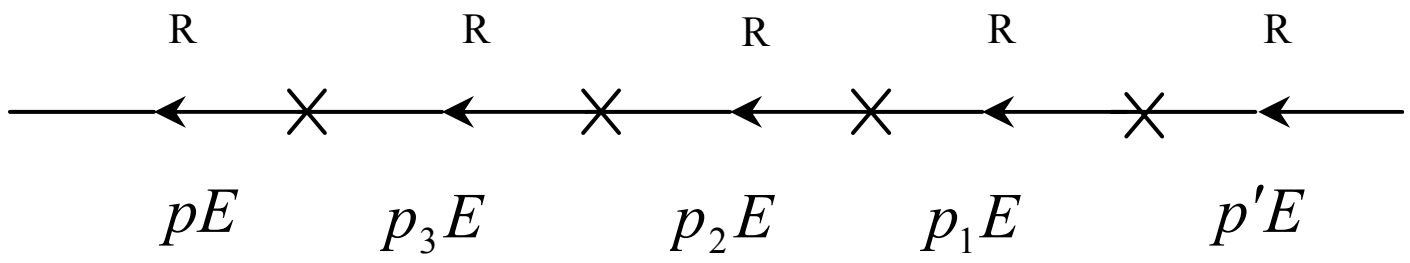

(50)

In the same way as the previous cases after some calculations we have:

$G_+^{(4)}(p,E)^{i=j=k=l} =$

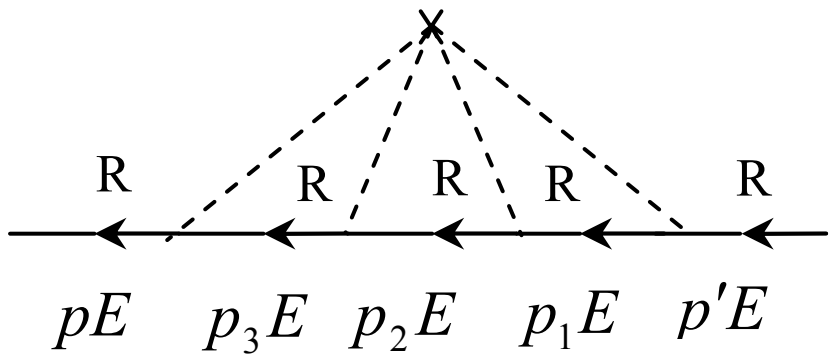

$$= n_i \int \frac{dp_3}{(2\pi\hbar)^3} \int \frac{dp_2}{(2\pi\hbar)^3} \int \frac{dp_1}{(2\pi\hbar)^3} V_{imp}(p - p_3) V_{imp}(p_3 - p_2) V_{imp}(p_2 - p_1) V_{imp}(p_1 - p)$$
$$G_+^{(0)}(p_3, E) G_+^{(0)}(p_2, E) G_+^{(0)}(p_1, E) \left[G_+^{(0)}(p,E)\right]^2 . \tag{51}$$

We also get fourth-order terms corresponding to diagrams with dangling impurity lines. For example the terms where the scattering is off different impurities $i \neq j \neq k \neq l$,

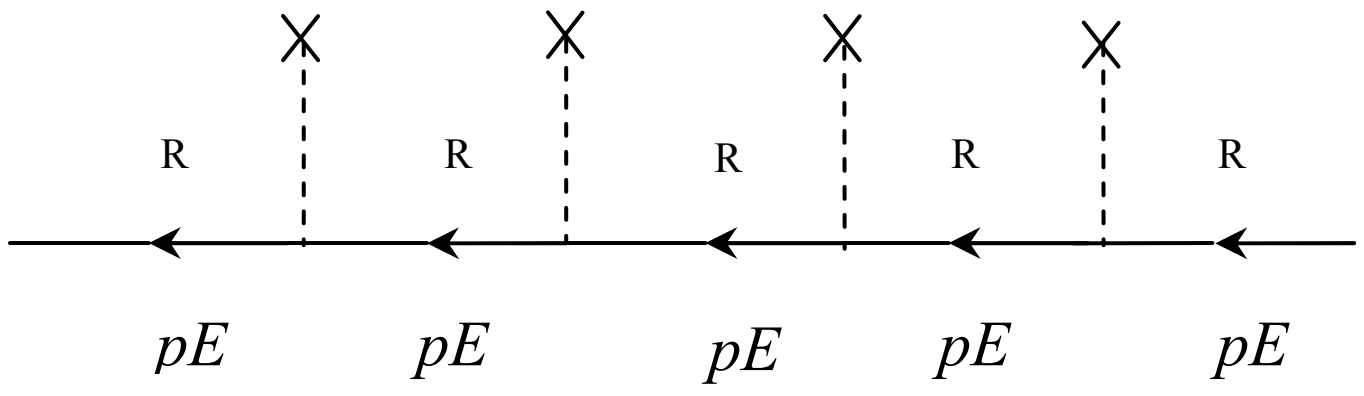

(52)

Or terms with two dangling lines, for example the term where $i \neq j = l \neq k$,

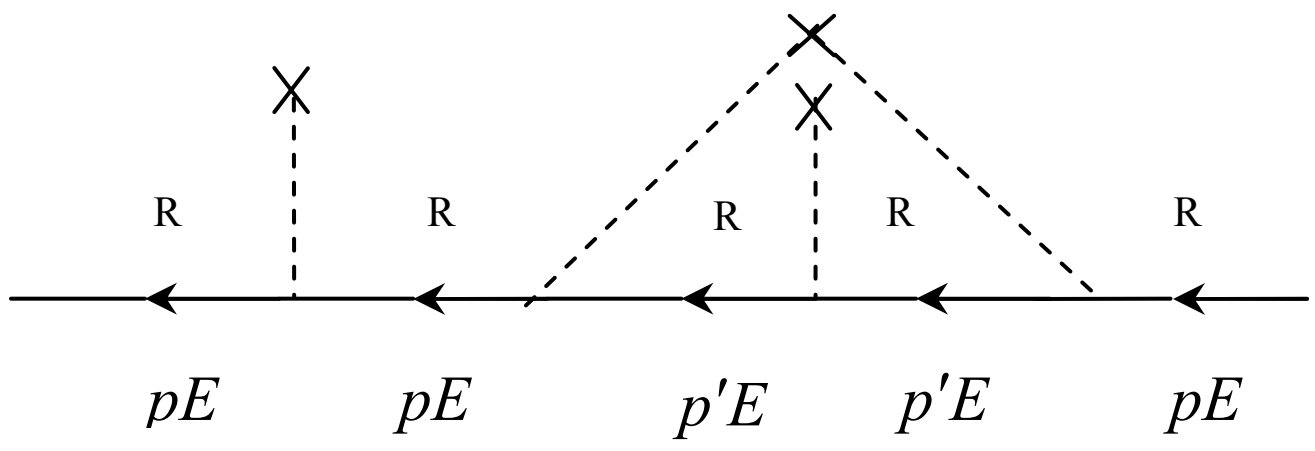

(53)

The contribution of such diagrams to $\delta\sigma^{(4)}(B)$ can be renormalized to zero. Nonvanishing contribution arises from double scattering off two different impurities For example the case where we consider $i = j \neq k = l$. The impurity phase factor is then

$$\left\langle e^{-\frac{i}{\hbar}(p-p_2)\cdot r_i} e^{-\frac{i}{\hbar}(p_2-p')\cdot r_k} \right\rangle = \delta_{p,p_2}\delta_{p_2,p'} = \delta_{p,p_2}\delta_{p,p'} = \delta_{p_3+p_2,p_3+p}\delta_{p_2+p_1,p'+p_1}. \tag{54}$$

Double scattering off two impurities can occur in three different ways: $i = j \neq k = l$, and the other two possibilities $i = k \neq j = l$ and $i = l \neq j = k$, they correspond to the following diagrams:

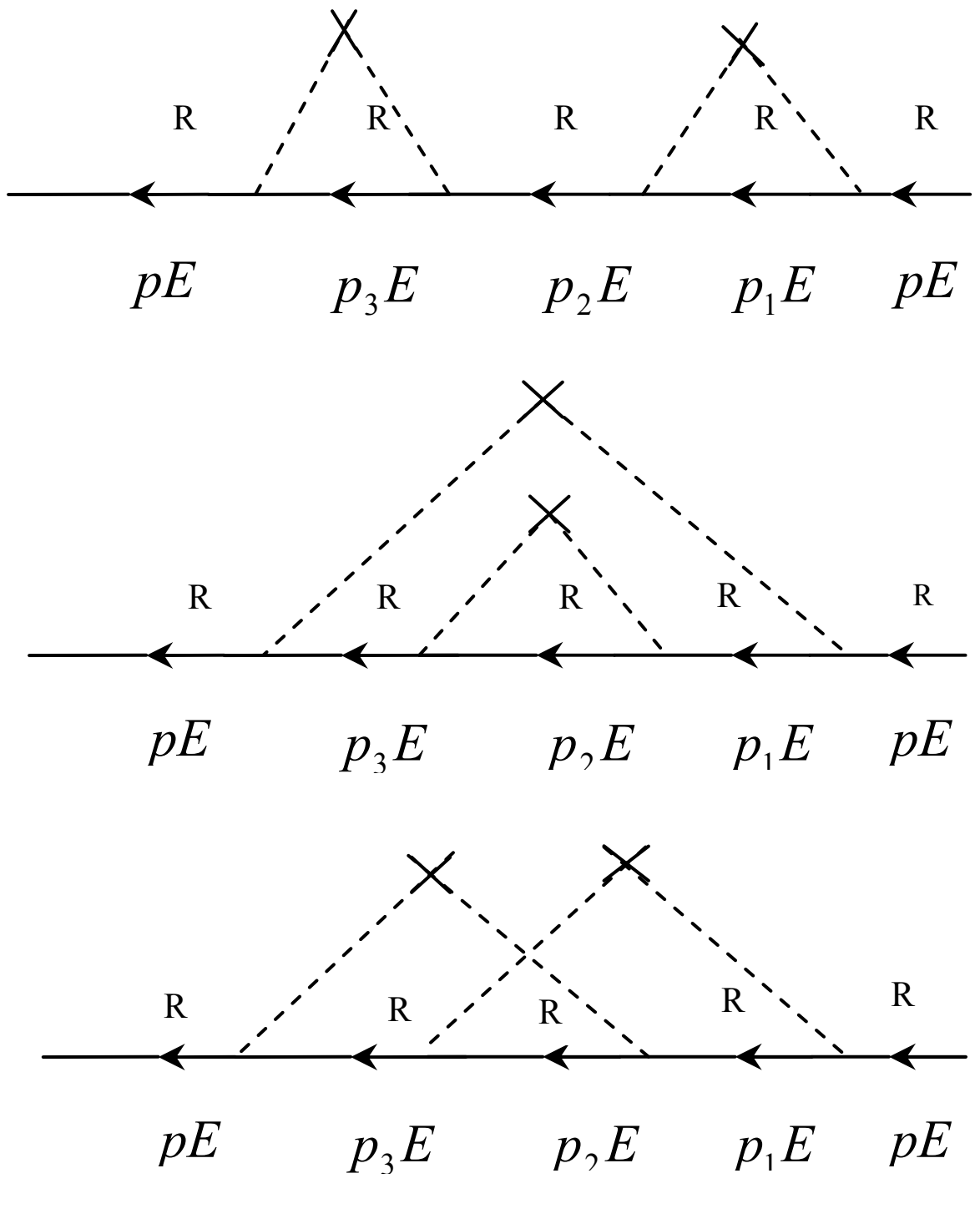

(55)

So the Feynman rules corresponding to the fourth order Feynman diagrams are as follows :

$$\delta\sigma^{(4)}(B) = -\sqrt{\frac{2}{\pi}} G_0 e^{\frac{1}{N_0}} \alpha \, Tr\left[G_{+}^{(4)}(p,E)^{i=j\neq k=l}\right],$$
$$-\sqrt{\frac{2}{\pi}} G_0 e^{\frac{1}{N_0}} \alpha \, Tr\left[G_{+}^{(4)}(p,E)^{i=k\neq j=l}\right], \qquad (56)$$
$$-\sqrt{\frac{2}{\pi}} G_0 e^{\frac{1}{N_0}} \alpha \, Tr\left[G_{+}^{(4)}(p,E)^{i=l\neq j=k}\right].$$

Therefore we get the self energy as follow

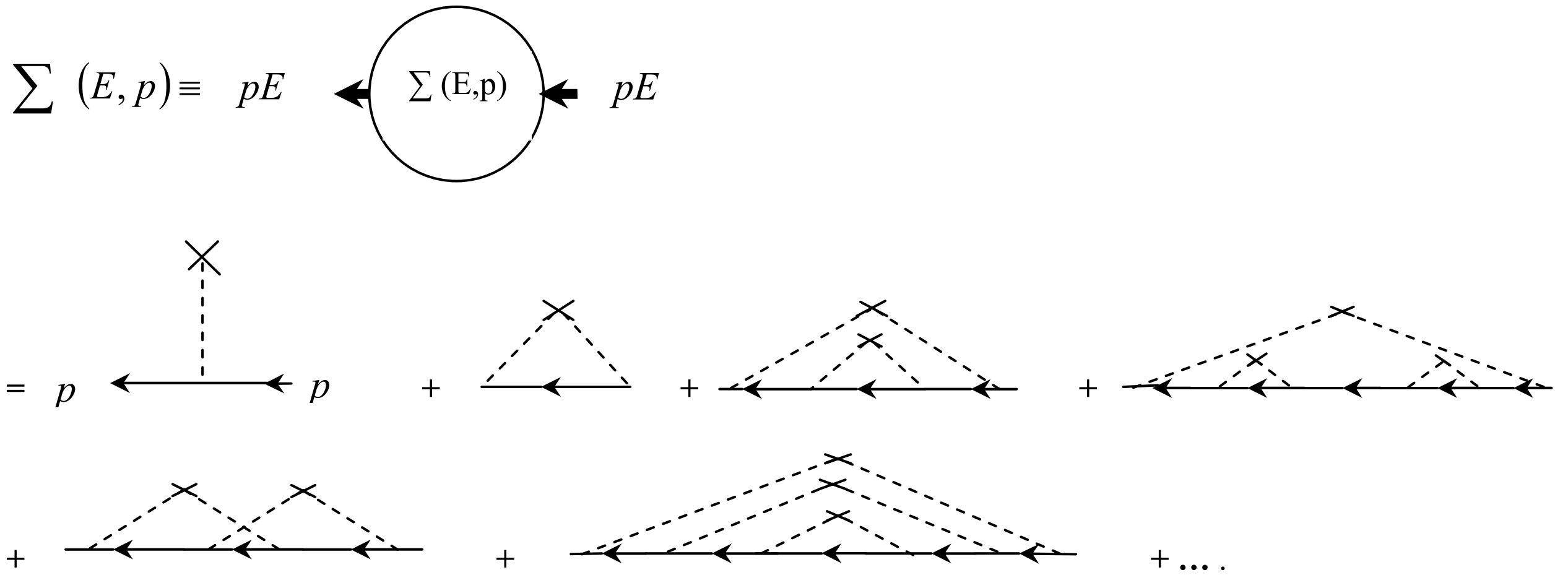

(57)

So we can construct

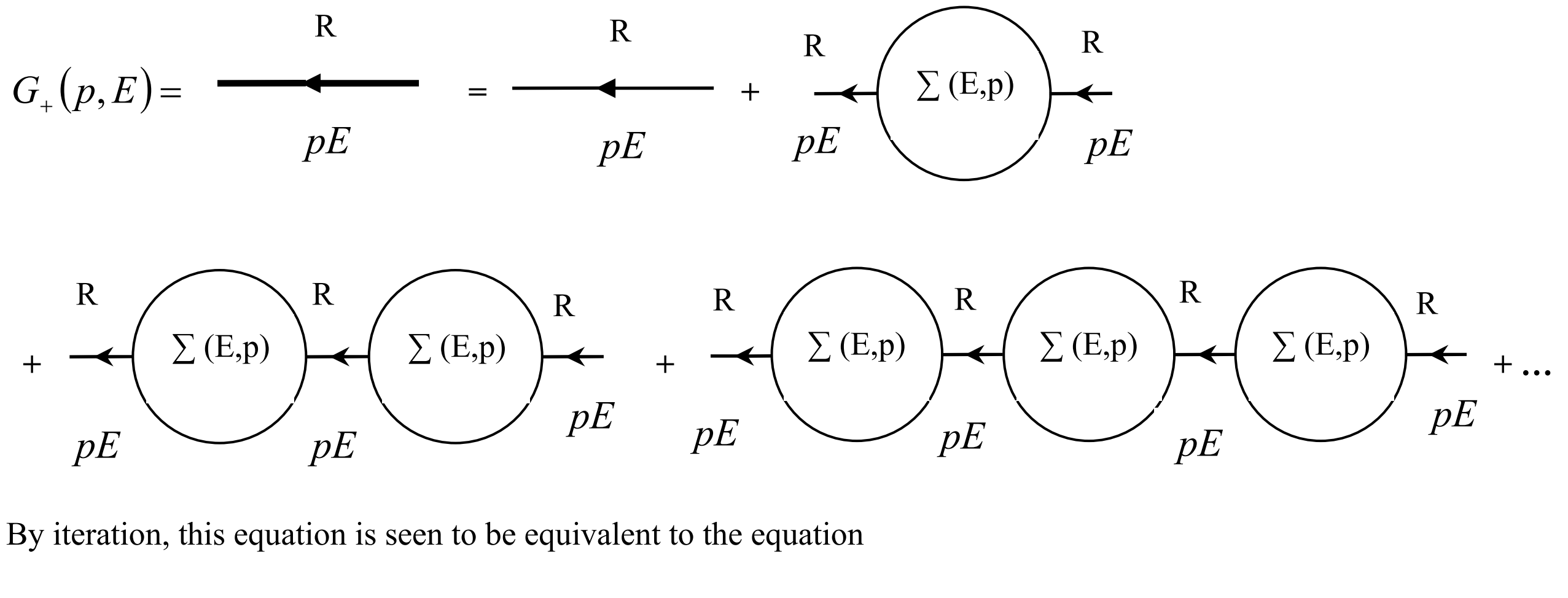

(58)

By iteration, this equation is seen to be equivalent to the equation

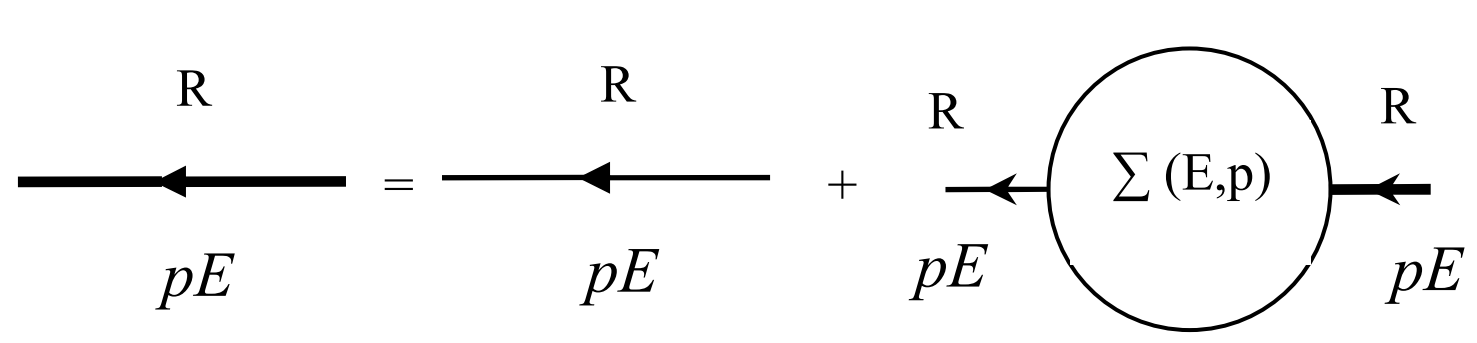

(59)

It is worth mentioning again that the same method can be used to derive the Feynman diagrams for the advanced Green function $G_-^{(0)}$.

Some representative samples of the kind of diagrams which give the correction to the retarded and advanced Green's function, e.g. $G_+^{(0)}VG_+^{(0)}$ , $G_-^{(0)}VG_-^{(0)}$ , … are depicted in fig. (4).

***b) Feynman diagrams for two-particles correlation functions***

Eq. (24) also involves two-particles correlation functions, e.g. $G_+^{(0)}VG_-^{(0)}$, $G_-^{(0)}VG_+^{(0)}$, … .

With the same method presented in section "a" one can find the Feynman diagrams for two-particles correlation functions.

Some few of such diagrams are depicted in fig. (5).

So in summary the quantum corrections to conductivity can be decomposed as a sum of infinite terms corresponding to different orders of quantum corrections to conductivity as illustrated in fig. (6).

Here a comment on what is the benefit to interpret the quantum corrections to conductivity (QCC) in terms of the Feynman diagrams, is in order. Feynman diagrams are graphical ways to represent interactions. They provide a convenient shorthand for the calculations. It has been shown in quantum electrodynamics (QED) that the quantum corrections on the mass and the charge of the electron can be expressed in terms of infinite numbers of Feynman diagrams. In this section we have shown that quantum corrections to conductivity of electrons as given by exact formula "(2)" can also be expressed in terms of infinite numbers of Feynman diagrams. As mentioned before this method provide a convenient shorthand for the understanding and calculating of QCC. They are a code we use to talk about QCC and provide a nice parallel between QCC and QED.

## CONCLUSION

Semiconductors are tremendously important in physics and electronic technology and play a key role in our daily lives. They are used in everything from greeting cards to satellites. In order to sustain these advancements, however we need to develop theoretical basis as well as experimental techniques of semiconductors. In this paper we present a new equation for quantum corrections to conductivity for semiconductors with various structures. At low temperatures and / or high disorders the Hikami et.al., [2,3] expression is not in good agreement with experiments even in low magnetic field for which the diffusion approximation is valid but Alavi-Rouhani expression in both high and low temperatures and/or high disordered is in good agreement with experiment. On the other hand from theoretical point of view Alavi-Rouhani expression establishes a powerful and rich basis for study of quantum corrections to conductivity. It relates the quantum corrections to the conductivity to the trace of Green function made by generator of $su_q(2)$ algebra, so the quantum corrections to the conductivity can be expressed as a sum of an infinite number of Feynman diagrams. The results presented here may be of interest for physicists working in the following areas of physics:

1. Condensed matter theory and the physics of semiconductors ( also electronic engineering ).
2. Low temperature physics .
3. mathematical physics- the application of quantum groups in physics.

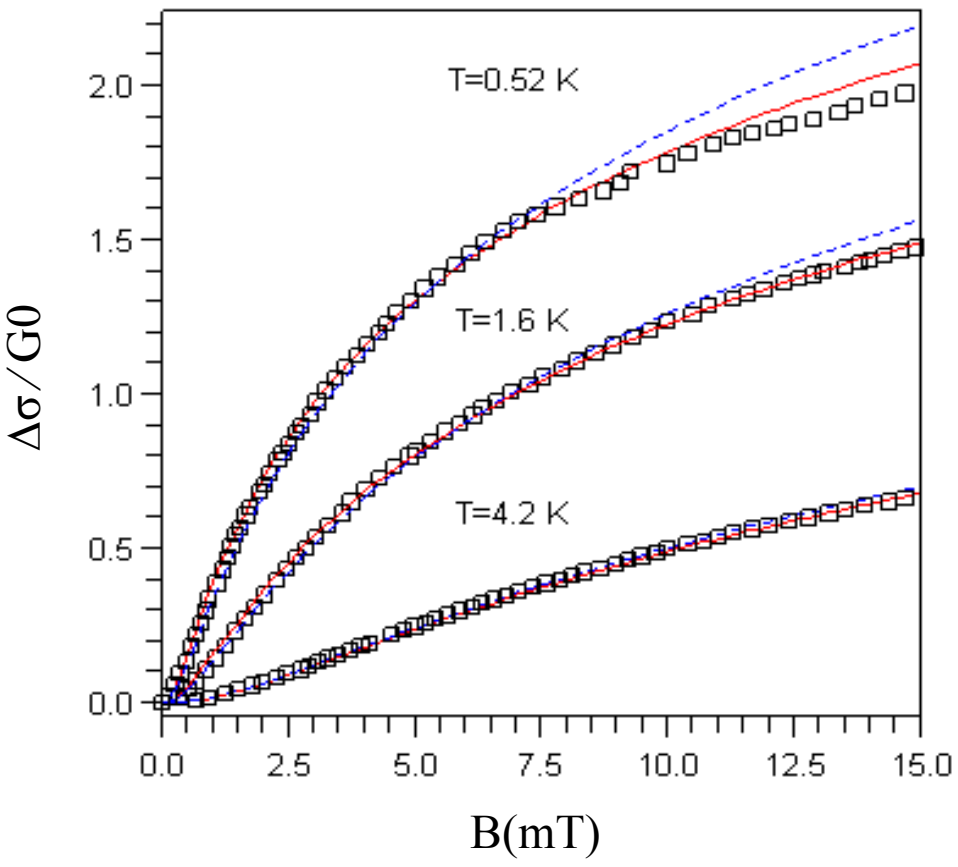

Figure 1. Magnetic field dependence of $\Delta\sigma / G_0$ for three various temperatures. Solid lines are the results of fitting by " (2)". Dashed lines are the results of fitting by "(1)". Open squares are experimental data. $B_{tr} = 0.03T$ , $\alpha = 1$ , $\beta = 203$ , 89 , 24 .9 and $a = 0.93, 0.89, 0.78$ , $\gamma = 0.007, 0.0135, 0.412$ for the temperatures $T = 0.52K, 1.6K, 4.2K$ respectively.

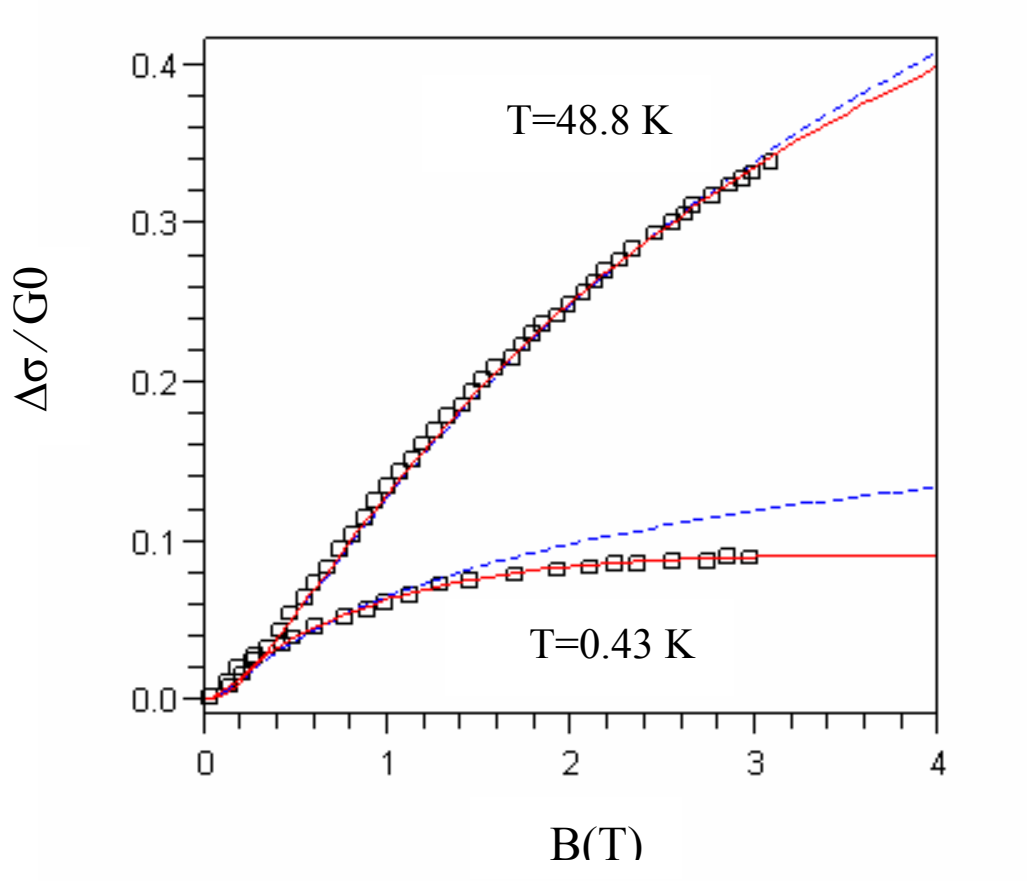

Figure 2. Magnetic field dependence of $\Delta\sigma / G_0$ for temperatures $T = 48.8K, 0.43K$. Solid lines are the results of fitting by " (2)". Dashed lines are the results of fitting by " (1)". Open squares are experimental data., $\alpha = 0.0893$ , $\beta = 7.81$ , $a = 0.059$, $\gamma = 0.0114$ for $T = 0.43K$ and $\alpha = 0.47$, $\beta = 12.1$, $a = 0.332$, $\gamma = 0.038$ for $T = 48.8K$ respectively. In both cases $B_{tr} = 5.3T$.

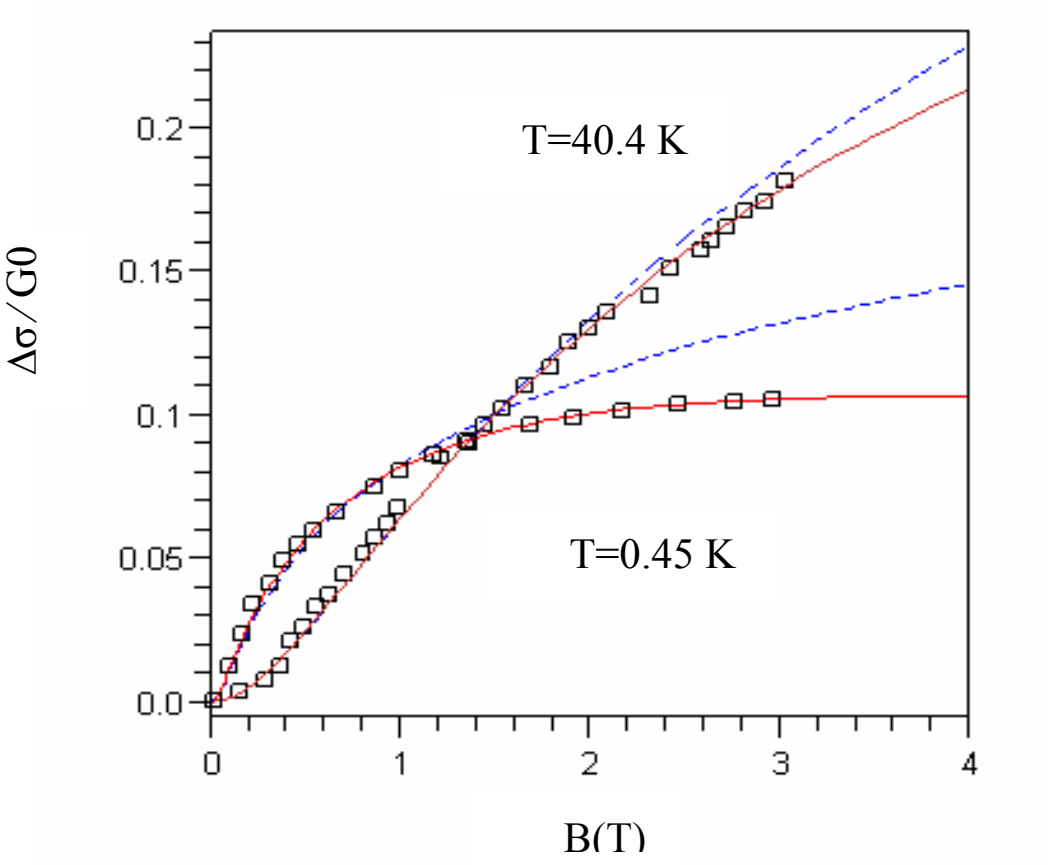

Figure 3. Magnetic field dependence of $\Delta\sigma / G_0$ for temperatures $T = 40.4K, 0.45K$. Solid lines are the results of fitting by “(2)”. Dashed lines are the results of fitting by “(1)”. Open squares are experimental data., $\alpha = 0.073$, $\beta = 14.95$, $a = 0.0497$, $\gamma = 0.0048$ for $T = 0.45K$ and $\alpha = 0.309, \beta = 7.65$, $a = 0.211, \gamma = 0.039$ for $T = 40.4K$ respectively. In both cases $B_{tr} = 6.3T$.

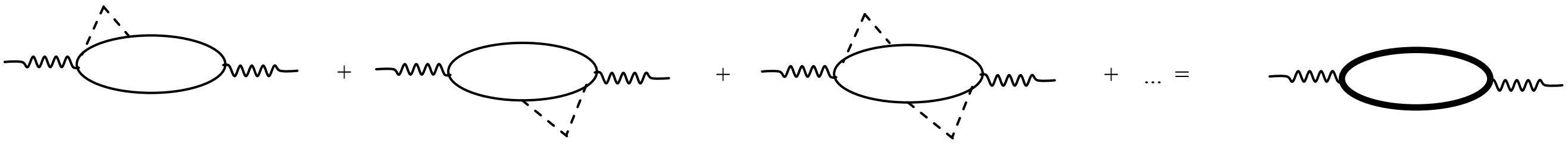

Fig. (4). Illustration of conductivity with the help of Feynman diagrams. Corrections to conductivity due to correction to the single-particle Green’s functions.

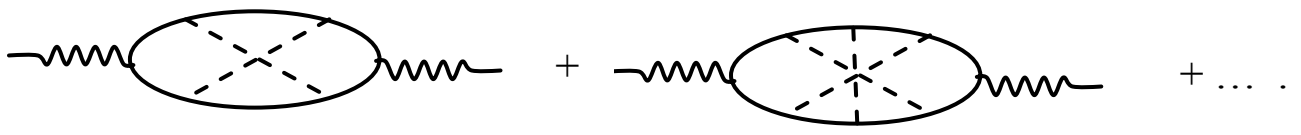

Fig. (5). Illustration of conductivity with the help of Feynman diagrams. Corrections to conductivity due to correction to the two-particles correlation functions.

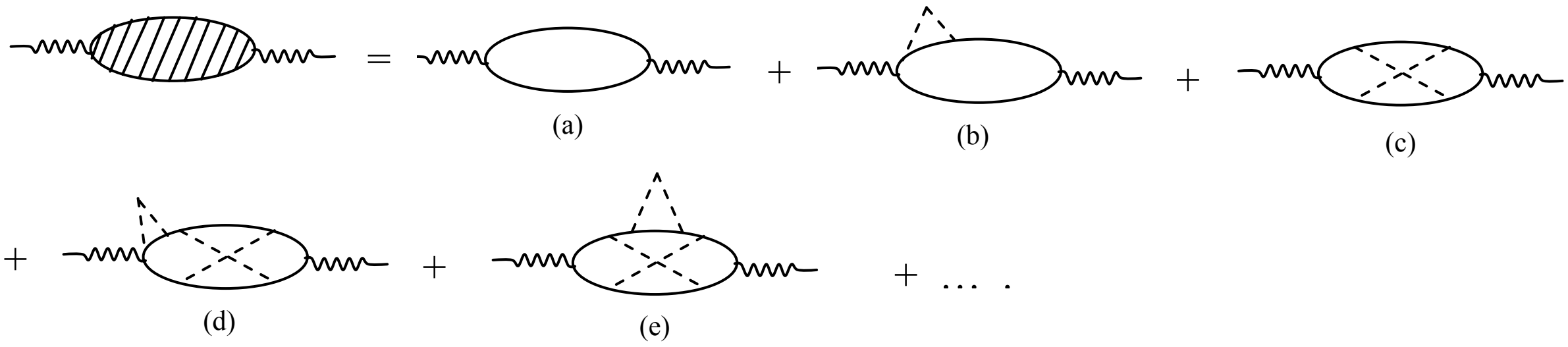

Figure. (6). Illustration of the conductivity with the help of Feynman diagrams. The total conductivity (hatched diagram), expressed like an infinite sum of diagrams involving the non-disordered Green's function $G^{(0)}$ (thin curve line) can be rewrite like an infinite sum of diagrams. The dashed lines refer to the potential.